\documentstyle[12pt]{article}
\title{}
\author{}
\begin{document}

\baselineskip=8mm plus 1mm minus 1mm

\begin{center}
{\bf The Forming of the Cosmological Constant}

\vspace*{.5cm}

V.~Burdyuzha${}^{1,3}$, O.~Lalakulich${}^{2}$, Yu.~Ponomarev${}^{3}$, G.~Vereshkov${}^{2}$
\end{center}

\vspace*{.3cm}

\noindent
${}^{1}$ Institut d'Astrophysique de Paris, 98 bis,
boulevard Arago, 75014 Paris, France \\
${}^{2}$ Rostov State University,
Stachki str. 194, 344104 Rostov on Don, Russia\\
${}^{3}$ Astro Space Center of Lebedev Physical
Institute of Russian Academy of Sciences,  \\
\hspace*{.3cm}Profsoyuznaya str. 84/32, 117810 Moscow, Russia\\

\vspace*{2cm}
\begin {abstract}
The problem of the physical nature and the cosmological constant genesis
is discussed.  This problem can't be solved in
terms of the current quantum field theory which operates with
Higgs and nonperturbative vacuum condensates and takes into account
the changes of these condensates during relativistic phase transitions. The
problem can't be completely solved also in terms of the conventional global
quantum theory: Wheeler-DeWitt quantum geometrodynamics does not
describe the evolution of the Universe in time (RPT in particular).
We have investigated this problem in the context of energies density of
different vacuum subsystems characteristic scales of which pervaid all energetic
scale of the Universe. At first the phemenological solution of the cosmological
constant problem and then the hypothesis about the possible structure of a
new global quantum theory are proposed.
The main feature of this theory is the irreversible evolution of geometry
and vacuum condensates in time in the regime of their selforganization.
The transformation of the cosmological constant in dynamical variable
is inevitably.
\end {abstract}
\newpage

     The cosmological constant problem is one of intriguing problems of
modern physics and astronomy. The suggestions for its solution have
attracted a lot of attention [1]. The idea of compensation of a initial vacuum
energy by vacuum condensates of quantum fields with cooling of cosmological
plasma during relativistic phase transitions is discussed (the short version
of this article using Zel'dovich's approximation was published in [2]).
The list of other adjustment mechanisms for cosmological constant can also
find in [1]. The terms of cosmological constant ($\Lambda$-term) and vacuum
energy are used practically synonymously in modern cosmology.

Today the strictly established fact is the physical vacuum is a complex
heterogenic system of classic and quantum fields consisting of three
subsystems:
1) zeroth weakly correlated vibrations of quantum fields;
2) zeroth strongly correlated vibrations of quantum fields producing
nonperturbative vacuum condensates;
3) quasiclassic (quasihomogeneous and quasistationary) fields usual
named Higgs condensates.
All these subsystems are included in Standard Model (SM). The
existence of zeroth weakly correlated vibrations has the
experimental confirmation: anomalous magnetic moment of electron, Lamb's
shift, Casimir effect, radiative corrections.  The values of
nonperturbative condensates which have firstly introduced  in [3] are usually
established in physics of vector mesons.  The question of Higgs condensate
existence will be decided after detection of Higgs bosons. Here we are
discussing  the cosmological constant problem in context of energy densities
of different vacuum subsystems.
The problem is that each vacuum
subsystem has a huge energy density, however the total value of vacuum
energy in the Universe today is near zero as observational data confirm
it [4]. Thus the phenomenon of selforganization of vacuum is
evident although the mechanism of selforganization of nonperturbative
condensates does not understand till now well (the possible models of
"nullification" of vacuum energy have been discussed also in [5-6]).

  Here selforganization of vacuum is the ability of a system to react
on outer condition such way to conserve itself local stability and
to evolve subsequently. Other aspect is that vacuum is a strongly
nonlinear system the intensity interactions in which depends on
outer conditions. The third aspect is the value $\Lambda \approx 0$
sure to be cosmologically preferable. The Universe with a large negative
$\Lambda$-term never become macroscopic, if the value of $\Lambda$-term is
large and positive the production of complex nuclear, chemical, biological
and cosmogonical structures is impossible. Our real Universe with the
observed structure hierarchy can exist when $\Lambda \sim 0$ only.

The energy density of vacuum in General Relativity is described by
constant $\Lambda$-term when the interactions of the vacuum subsystems
with matter is negligible:
$$R_{\mu \nu} - \frac{1}{2} g_{\mu \nu} R = 8 \pi G T_{\mu \nu} +
\Lambda g_{\mu \nu} . \eqno(1)$$
The problem is to calculate this constant in the modern and previous epochs.
Today we can more exactly define the physical meaning of $\Lambda$-term
which must contain the energy-momentum tensor (EMT) of gravitational
vacuum $T_{\mu \nu(g)}$ and EMT of quantum fields $T_{\mu \nu(QF)}$:
$$R_{\mu \nu} - \frac{1}{2} g_{\mu \nu} R = \mbox{\ae} (T_{\mu \nu (g)} +
<T_{\mu \nu(QF)}>) = \mbox{\ae} (g_{\mu \nu} \Lambda_{g} + g_{\mu \nu}
\Lambda_{QF}). \eqno(2)$$
Here $\mbox{\ae} = (10^{19} \; Gev)^{-2}$ is the gravitational constant in
the system of units where $\hbar = c = 1; <T_{\mu \nu(QF)} >$ is EMT of
quantum field averaged on some martix of density, which contains the
information about state of plasma and vacuum of elementary particles.
For $\mid R^{\nu}_{\mu} \mid \ll \mbox{\ae}^{-1}$ the averaged EMT of quantum
fields is:
$$
<T_{\mu \nu(QF)} > = < 0 \mid T_{\mu \nu(QF)} \mid 0 > = g_{\mu \nu} \Lambda_{QF};
$$
$$
T_{\mu \nu(g)} = g_{\mu \nu} \Lambda_{g}.
$$
Here $\Lambda_{g}$ is the second fundamental constant of the gravitation
theory accounting a gravitational vacuum condensate. Other words
$\Lambda$-term must contain two items:
$$
\Lambda = \Lambda_{g} + \Lambda_{QF} \eqno(3)
$$
which have practically exactly compensated each other since the
observable value of $\Lambda$-term (the cosmological constant) is near
zero [4]. Many theorists find that $\Lambda$-term must be calculated
in a unified theory of all interactions and the separation on two items
is artificial (naturally it is so). But the subject of our research is a
heterogenic system (geometry + vacuum + fields of matter) and for a
arbitrary state of this system is not possible to extract the vacuum
energy as the separate item in $<T_{\mu \nu(QF)} >$. The constant
$\Lambda_{QF}$ can arise as a physical magnitude if two conditions are
carried out: 1) a vacuum subsystem after relaxation must reach the
equilibrium state with plasma of elementary particles; 2) the
temperature and density of plasma must be small in comparison with
critical values of magnitudes which characterize the point of a
relativistic phase transition (RPT) (more detail see [7]). For illustration
we shall use the simplest chain of RPT which may take place during
initial evolution in our Universe:

\vspace*{.3cm}
\begin{tabular}{ccccc}
P & $\Longrightarrow$ & $D_{4} \times [SU(5)]_{SUSY}$ & $\Longrightarrow$ &
$D_{4} \times [U(1) \times SU(2) \times$ \\
  & $10^{19} Gev$     &                               & $10^{16} Gev$ &  \\
\end{tabular}

\vspace*{.3cm}

\begin{tabular}{ccc}
$\times SU(3)]_{SUSY}$ & $\Longrightarrow$ & $D_{4} \times U(1) \times SU(2)
 \times SU(3)$ \\
                       & $10^{5}-10^{10} Gev$ &  \\
\end{tabular}

\vspace*{.3cm}

\begin{tabular}{cccc}
$\Longrightarrow$ & $D_{4} \times U(1) \times SU(3)$ & $\Longrightarrow$ &
$D_{4}\times U(1)$ \\
 $10^{2} Gev $        &                                  & $150 Mev$ &\\
\end{tabular}

Of course whole chain is our proposal but one can be sure only in two
last transitions: electroweak (EW PT) and quark-hadron PT. EW PT occurs
at temperature about $10^{2}$ GeV and it is accompanied by appearence of a
Higgs condensate decreasing vacuum energy.
In the interval of temperatures $150 \; Mev < T < 100 \; Gev$ the vacuum
in the Universe was in the state of spontaneously breaking $SU(2)$
symmetry (for these temperatures the quark-gluon subsystem was in the
state of deconfinement that is the quark-gluon vacuum condensate was
absent).  The value of Higgs condensate is negative and it have been
calculated many times in different models. We use the expression:
$$\Lambda_{SM} = - \frac{m^{2}_{H} m^{2}_{w}}{2g^{2}} - \frac{1}{128
\pi^{2}} (m^{4}_{H} + 3 m^{4}_{z} + 6 m^{4}_{w} - 12 m^{4}_{t}).
\eqno(4)$$
Here the first term is the energy density of a quasiclassical Higgs
condensate. The second term is the polarization of vacuum by quantum fields.
Excepting $t$-quark others fermions are very light and they involve a
negligible small contribution in formula (4). Boson contributions are
negative but fermion ones are positive. The numerical values of
all constants except for Higgs boson mass are known from experiments
(see [8]). The limitations on a Higgs boson mass can be found from the
condition of vacuum stability:
$$x^{2} + x (\frac{1}{2a} - \frac{4ab}{9}) - \frac{2b}{3} > 0,$$
$$x < \frac{1}{a} + \frac{4ab}{9}, \eqno(5)$$
here $x = m^{2}_{H}/m^{2}_{w} ; a = 3g^{2}/128 \pi^{2}; b =
\frac{12m^{4}_{t} - 3m^{4}_{z} - 6m^{4}_{w}}{m^{4}}, g^{2} = 0.43$ is
the gauge constant $SU_{L} (2)$ group; $m_{w} = 80 \; Gev, m_{z} = 91
\; Gev, m_{t} = 175 \; Gev$.
Inequalities (5) give the interval of possible values of still no
experimentally detected  Higgs boson $36 \; GeV < m_{H} < 2480 \; GeV$
that does not contradict modern experimental restrictions $m_{H} > 75
\; GeV$.
Substituting these values of $m_{H}$ into (4) one can see that the mutual
compensation of positive and negative contributions in vacuum density energy
in SM is prohibited by the condition of stability. The last
conclusion has general character. It is an important moment of our
consideration. For decreasing of any symmetry during RPT the vacuum energy
must decrease.
So in order to have $\Lambda \approx 0$ one have to introduce ad hoc
the initial positive $\Lambda$-term. Then the decreasing of vacuum
energy during RPT can be considered as the compensation of an initial
positive value.
It seems to be the phenomenological solution of the $\Lambda$-term problem.

The next question arises immediately. Can zeroth value of $\Lambda$-term be
obtained as the consequence of the inner structure of a theory? The answer was
searching in terms of SUSY theories. The idea is based on the main
feature of such theories: the general contributions from Higgs
condensates are nearly compensated. The residual part must be compensated by
radiative corrections. This has been made in the multidimentional superstring
model after the special compactification in one-loop approximation. The
possibility of such coordination is prompted by mathematical formalism of
strongly nonlinear theory in which the state with $\Lambda = 0$ has the
status of a special branch of nonlinear equations. Quite evidently that the
coordination of vacuum subsystem states was realized during cosmological
evolution that is here we have all indicators of vacuum selforganization. It
is pertinently to recall the anthropic principle (probably the
selforganization of vacuum has provided the life of organic type in
the Universe).
We want to stop also more detail on QCD nonperturbative vacuum since the
extrapolation of QCD ideas to more deep structure levels of matter [9]
and to quantum gravity scales [10] is almost inevitably. A
nonperturbative quark-gluon condensate as the element of the theory is
included in SM . Without representations about this condensate the
confinement phenomena of quarks and gluons is not possible to
understand. The investigation of QCD equations has shown that
the confinement phenomena takes place if vacuum correlatories of quark-
gluon fields is not zero:
$$<0 \mid G_{\mu \nu} G^{\mu} \mid 0 > \;\; > 0; \;\;\;\; <0 \mid
\bar{q} q \mid 0 > \;\; < 0  \eqno(6)$$
(here for simplisity we are limiting the discussion of quantum
correlatories which are quadratic from fields but certainly in the
nonperturbative vacuum correlatories of any order are not zero). In the
perturbative vacuum these values after renormalization equal zero.
Inequalities (6) have the status of rigorous theoretical results
however they say nothing about the space-time dynamics of nonperturbative
fluctuations forming of a quark-gluon condensate.Fortunately the quantitative
level of nonperturbative fluctuations can be established from an experiment
the reply about their nature has not even. Next values of condensates
have been found [3]:
$$< 0 \mid \frac{\alpha_{s}}{\pi} \; G_{\mu \nu} G^{\mu \nu} \mid 0 > =
(360 \pm 20 \; MeV)^{4} \approx 27 \lambda^{4}_{QCD} \eqno(7)$$
$$< 0 \mid \bar{u} u\mid 0 > \approx < 0 \mid \bar{d} d \mid 0 >
\approx < 0 \mid \bar{s} s \mid > \approx (-225 \pm 25 \; MeV)^{3} \approx
 -2.8 \lambda^{3}_{QCD}$$
and then the energy density of nonperturbative QCD vacuum is:
$$\epsilon_{vac} = - \frac{9}{32} < 0 \mid \frac{\alpha_{s}}{\pi} \;
G_{\mu \nu} G^{\mu \nu} \mid 0 > + \frac{1}{4} <0 \mid m_{u} \bar{u}·
u\mid 0 > +$$
$$+ < 0 \mid m_{d} \bar{d} d \mid 0 > + < 0 \mid m_{s} \bar{s} s \mid 0
> = - 8.2 \lambda^{4}_{QCD}. \eqno(8)$$
Here $m_{u} = 4.2 \; Mev; m_{d} = 7.5 \; Mev; m_{s} = 150 \; Mev$ are
masses of light quarks satisfacting to the condition $m_{q} \le
\lambda_{QCD}; \lambda_{QCD} = 160 \; Mev$. What is the physical nature
of nonperturbative fluctuations forming a quark-gluon condensate?
Probably the nonperturbative vacuum is a bose-condensate of dions and
antidions. The sum charges and averaged on large distances
gluon (chromoelectrical and chromomagnetic) fields in vacuum equal of
course zero however fluctuations of these fields in scales of the correlated
lenght of a dion condensate are not zero. The average  values of square
of fluctuating gluon fields is the basic characteristics of
nonperturbative QCD-vacuum. Fluctuations of quark fields are probably
induced by fluctuations of gluon fields. This follows from the relation:
$$<0 \mid \bar{q} q \mid 0 > = - \frac{1}{12 \mu_{q}} < 0 \mid
\frac{\alpha_{s}}{\pi} \; G_{\mu \nu} G^{\mu \nu} \mid 0 >, \eqno(9)$$
here
$$
 \mu_{q} = \left\{
\begin{array}{rl}
\lambda_{QCD}: q = u, d,s  & m_{q} \le \lambda_{QCD}\\
m_{q} : q = c,b, t  & m_{q} \gg \lambda_{QCD}
\end{array}.\right. $$
Thus for $T < \lambda_{QCD} = 160 \; Mev$ a nonperturbative quark-gluon
vacuum is the state of a dion condensate with negative energy density
(the classic prototype of dions is nonlinear solutions of Yang-Mills
equations similar to solitons, instantons, monopoles). That is today
vacuum in the Universe has the confinement phase and the modern value
of $\Lambda$-term can be calculated using formula:
$$\Lambda_{QF} = \Lambda_{SM} + \epsilon_{vac}. \eqno(10)$$
Here $\epsilon_{vac}$ is the energy density of quark-gluon vacuum (see
(8)) and $\Lambda_{SM}$ is the constant taking phenomenologically into
account all vacuum structures on energy scales more than
$\lambda_{QCD}$.
It can be easily understood that only $\Lambda_{SM}$ coming from Higgs
vacuum can be compensated by SUSY mechanism. The similar mechanism for
$\epsilon_{vac}$ coming from nonperturbative vacuum is absent until
the problem of dimensional transmutation will not be solved. Finally our
conclusion is the problem of $\Lambda$-term can't be solved in terms of
the current field theory.

It is  worthwhile to say some  words about calculations
of today value of $\Lambda$-term which were carried out recently in [2]
using Zeldovich's approximation. The vacuum condensates (Higgs and
nonperturbative one) in the modern quantum theory are macroscopic mediums
with quasiclassical properties. The periodic collective motions in these
mediums are perceived as pseudogoldstone bosons. For temperature  of
chiral symmetry breaking ($T_{c} \sim 150 \; Mev$)
the main contribution in periodic collective motions of a
nonperturbative vacuum quark-gluon condensate introduces $\pi$-mesons
as the lightest pseudogoldstone particles. That is here in essence the
spectrum of excitations reflects the properties
of a ground state.  Ya.Zeldovich [11] attempts to account for a nonzero
vacuum energy density of the Universe in terms of quantum fluctuations
(the gravitational force between particles in the vacuum fluctuations
as a higher-order effect) inserting in the finded them formula $\Lambda
= 8 \pi G^{2} m^{6} \hbar^{-4}$ the mass of proton or electron.
Calculations have shown that the agreement of the result with the observed
value is not good. Kardashev [12] had proposed to modify Zeldovich's formula
and to use mass of pions:
        $$\Lambda= 8 \pi G^{2} m^{6}_{\pi} h^{-4}. \eqno(11)$$
(note, here we have  $h$ instead of $\hbar$). Remarkably that the calculated values of
$\Lambda$-term using Zeldovich's formula gives $\Omega_{\Lambda} = 0.7$
if $H_{o} = 72.5 \; (km/s)/Mps$ (here $H_{o}$ is the Hubble constant)
and $\Omega_{\Lambda} = 0.8$ if $H_{o} = 67.8 \; (km/s)/Mps$ (see the
table in [2]).
The mistic agreement of formula (11) with the observable value
nevertheless gives no physical explanation of $\Lambda$-term problem and our
conclusion about the necessity to go beyond the current field theory does not
change.But it may be the first approximation to our understanding and
the calculation of today value of the cosmological constant. The next step can
be made by quantum geometrodynamics (QGD).

Initially two approaches have been proposed.
The first one is Hawking's idea to introduce in the theory (besides the usual
fields describing both vacuum and particles) some special fields which
concern to vacuum only. Such special fields were called 3-forms, 4-forms
and so on. As S.Hawking has shown [13] that the more probable state of the
Universe was the state with $\Lambda_{eff} = 0$ since
$$P (\Lambda_{eff}) \sim exp \; (\frac{3 \pi}{\mbox{\ae}^{2} \Lambda_{eff}}).
\eqno(12)$$
here $\Lambda_{eff}$ is the sum of usually discussed $\Lambda$-term and
 the contribution from 3-forms.
The problem is that is the nature of "formes" and how they can be
experimentally observed in a local experiment (beside the influence on
$\Lambda$-term).
A more deep step in the investigation of $\Lambda$-term problem was made by
S.Coleman [14] who took into attention the realistic effect of microscopic
quantum fluctuations of space-time topology at the Planck scale
(worm holes). In this approach the more probable the state of the Universe
has had a more sharp peak than S.Hawking's distribution :
$$P (\Lambda_{eff}) \sim exp \;(exp \; \frac{3 \pi}{\mbox{\ae}^{2}
\Lambda_{eff}}) \eqno(13)$$
Here $\Lambda_{eff} = \Lambda_{QF} + \Lambda_{WH}$, where $\Lambda_{WH}$
is the contribution of worm holes.
This approach allowes to give the unified conception of vacuum. Recall,
the QCD vacuum also is a system of quantum topological fluctuations.

Thus these are two limiting  points on the energetic scale of the Universe:
the first point is $\Lambda_{QCD} = 150 \; MeV$  which are experimental
fact described in experimentally tested theories; the second point is
fluctuations at the
Planck scale which are the direct consequence of Quantum Gravity. Both
types of fluctuations have a geometrical origin.
We propose the quantum topological fluctuations can exist at other
intermediate scales. This idea is realized in preon theories of
elementary particles where we have the hierarchy of nonperturbative
condensates instead of Higgs condensates[9]. This approach seems to be
favourable because it gives the unified  picture of vacuum. (recall the Higgs
bosons are still
experimentally undetected and Higgs conception of vacuum is not confirmed).

Certainly neither Hawking's approach not Coleman's approach do not
solve the $\Lambda$-term problem because their results were obtained in
terms of Wheeler-De Witt QGD which does not describe the quantum evolution
of the Universe. In the real Universe the energy of vacuum has changed in
time in the processes of the RPT (this was inevitably).

To solve this problem, a new version of QGD have to be formulated. The
new theory must theoretically describe the evolution of the Universe wave
function in time. The dynamical processes in vacuum and elementary
particle plasma which influence on wave function evolution should be
taken into account. As we think at least three steps should be made.

The first step is to change the status of $\Lambda$-term from the
cosmological constant to the dynamical variable. In classical theory it
was described by Weinberg in [6] who has rewritten the Einstein's equations
in the special gauge in the form no containing $\Lambda$-term. In his
theory $\Lambda$-term is an integral of motion. Having fixed
$\Lambda$-term one finds solutions of equations.

We propose at the second step one should do the same in quantum theory.
Here, in distinguishing from classical theory the integral of motion can't
have an arbitrary value.
The spectrum of the  allowed values is fixed by the eigenvalues of the
superhamiltonian. This spectrum can be discrete or continuous but from
our point of veiw near the small
values of $n$ it is discrete. This
superhamiltonian will describe interactions between
topological fluctuations of different scales. Any changes at one scale
lead to rebuilding of vacuum condensates at other scales. That is
we refer to as quantum selforganization of vacuum. The usual quantum theory
is the reversible theory. So at the third step it
is worth to recall the R.Penrose suggestion
that  quantum evolution is to be irreversible. This can be
realized, for example, by suppression of quantum transitions with $\mid
\Lambda \mid$ increasing.

Finally in short way the result is expressed by the formula:
$$\Lambda = \Lambda_{QF} + \Lambda_{WH} + \Lambda_{G}. \eqno(14)$$
Here $\Lambda_{QF}$ is formed by the zeroth vibrations of quantum
fields and by nonperturbative condensates; $\Lambda_{WH}$ is formed
by worm holes; the evolution of a gravitation vacuum condensate (GVC) forms
$\displaystyle \Lambda_{G} \equiv \frac{9\pi^{2}}{2 \mbox{\ae}^2}\lambda_n$
where $\lambda_n$ defines the spectrum of GVC possible states. The general
for all items in this formula is that they were created during evolution of
the Universe.

The value $\Lambda \approx 0$ (but no $\Lambda =  0$) can be explained
in the following way. According to mentioned above hypothesis about the
discrete spectrum the line with $\Lambda = 0$ must absent
but anyway there is a line with $\Lambda$ nearest to zero. During Universe
evolution the series of quantum transitions lead to this
state as the final state. The inverse transitions, according to R.Penrose,
must be suppressed. This is the process of vacuum  selforganization. The strategy
for the vanishing cosmological constant suggested recently by S.Adler [15] has
also included understanding of changes in the vacuum sector in the presence of
scale invariance breaking. Note,that the transformation of the cosmological constant
in dynamical variable has been already made by the introduction of quitessence [16]
althought it was an artificial step.

\vspace*{.3cm}

\noindent
{\bf References}

1. A.D.Dolgov, Proc. of the XXIV-th Rencontres de Moriond. Series:
Moriond Astrophysical Meeting, Les Arc, France, eds. J.Audouze and
J.Tran Thanh Van, p. 227 (1989); M.Gasperini, Phys. Lett. B224, 49
(1989); G.Lavrelashvili, V.Rubakov, P.Tinyakov, Proc. of 5th Seminar on
Quantum Gravity. Eds. Markov M.A., Beresin V.A., Frolov V.P. World
Scientific. Singapore, p.27 (1991); Y.Fujii, T.Nishioka Phys. Rev D.
42, 361 (1990); S.M.Carroll, W.H.Press and E.L.Turner, Ann. Rev. Astron.
Astrophys. 30, 499 (1992); M.M.Fukugita, T.Yanagida. Proc. of
RESCUE Symp. "The cosmological constant and the evolution of the
Universe". Eds. Sato K., Suginohara T., Sugiyama N. Universal Acad.
Press, Tokyo, p. 127 (1996); Efstathiou G., ibid, p. 225; Kolb E.M.,
ibid, p. 169; A.Vilenkin, ibid, p. 161; Martel H., Shapiro P.R.,
Weinberg S. astro-ph/9701099; S.Capozziello, R.de Rits, A.A.Marino.
Nuovo Cimento, 112B, 1351 (1997) and gr-qc/9806043; F.Pisano and
M.D.Tonasse, hep-ph/9701310; F.G.Alvarenga and N.A.Lemos. General Rel. and
Gravitation 30, 681 (1998); E.L.Guendelman and A.B.Kaganovich
gr-qc/9806053; N.Turok and S.Hawking, hep-th/9803156; G.M.Vereshkov,
Preprint of Lebedev Physical Inst. (1998); P.Ferreira, M.Joyce,
astro-ph/9711102; A.Starobinsky, Pis'ma ZhETF 68, 721 (1998); N.Straumann
Eur.J.Phys. 20, 419 (1999).

2. V.Burdyuzha, Proc. of the Sixth Intern.Symposium "Particles, Strings
and Cosmology" PASCOS-98 p.101 (1998); V.Burdyuzha, O.Lalakulich, Yu.Ponomarev
and G.Vereshkov, Preprint of Yukawa Inst. for Theoretical Physics (YITP-98-50)
and gr-qc/9907101.

3. M.A.Shifman, A.I.Vainstein, V.I.Zakharov  Nucl. Phys. B147, 385 (1978).

4. S. Perlmutter et al. Nature 391, 51 (1998), astro-ph/9812473 and Astrophys. J. 517, 565 (1999);
B.Schmidt et al., Astrophys. J. 507, 46, (1998).

5. A.D.Dolgov, Phys. Rev. D 55, 5881 (1997) and astro-ph/9708045.

6. S.Weinberg, Rev. Mod. Phys. 61, 1 (1989) and astro-ph/9610041.

7.  V.Burdyuzha, O.Lalakulich, Yu.Ponomarev, G.Vereshkov,
Phys. Rev. D 55, 7340R (1997) and astro-ph/9604124.

8. R.M.Barnett et al., Phys. Rev. D54 (1996).

9.  V.Burdyuzha, O.Lalakulich, Yu.Ponomarev, G.Vereshkov, Phys. Lett. B
(submitted) (1999); Preprint of Yukawa Inst. for Theoretical Physics
(YITP-98-51) and hep-ph/9907531.

10. G.Preparata, S.Rovelli, S.-S.Xue, gr-qc/9806044.

11. Ya.B.Zeldovich, Pis'ma JETP 6, 883 (1967).

12. N.S.Kardashev , Astron. Zh. 74, 803, (1998).

13. S.W.Hawking, Phys. Lett. B 134, 403 (1984).

14. S.Coleman, Nucl. Phys. B 310, 643 (1988).

15. S.L.Adler, General Rel. and Gravitation  29, 1357 (1997).

16. R.R. Caldwell, R.Dave, P.J.Steinhardt Phys.Rev.Lett. 80, 1582, (1998);
L.Wang, R.R.Caldwell, J.P.Ostriker, P.J.Steinhardt astro-ph/9901388.

\end{document}